\documentclass[conference]{IEEEtran}
\usepackage{amsmath,amssymb,amsfonts}
\usepackage{algorithm}
\usepackage{array}
\usepackage{textcomp}
\usepackage{stfloats}
\usepackage{url}
\usepackage{verbatim}
\usepackage{graphicx}
\usepackage{colortbl}
\usepackage{flushend}

\usepackage{tabularx}

\usepackage[style=ieee, backend=bibtex]{biblatex}
\usepackage{algpseudocode}
\usepackage{subcaption}
\usepackage{hhline}
\usepackage{multicol}
\usepackage{multirow}
\usepackage[T1]{fontenc}
\usepackage{orcidlink}

\usepackage{braket}

\usepackage{threeparttable}

\usepackage[utf8]{inputenc}
\usepackage{pgfplots}
\DeclareUnicodeCharacter{2212}{−}
\usepgfplotslibrary{groupplots,dateplot}
\usetikzlibrary{patterns,shapes.arrows}
\pgfplotsset{compat=newest}
\pgfplotsset{compat=1.11,
    /pgfplots/ybar legend/.style={
    /pgfplots/legend image code/.code={%
       \draw[##1,/tikz/.cd,yshift=-0.25em]
        (0cm,0cm) rectangle (15pt,0.8em);},
\hline   },
}

\begin{document}

\title{\fontsize{20pt}{21.0pt}\selectfont Peephole Optimization for Quantum Approximate Synthesis}

\author{\IEEEauthorblockN{Joseph Clark \orcidlink{0009-0000-0295-0083}}
\IEEEauthorblockA{\textit{Dept. of Electrical Engineering and Computer Science} \\
\textit{University of Tennessee}\\
Knoxville, USA \\
jclar168@vols.utk.edu}
\and
\IEEEauthorblockN{Himanshu Thapliyal \orcidlink{0000-0001-9157-4517}}
\IEEEauthorblockA{\textit{Dept. of Electrical Engineering and Computer Science} \\
\textit{University of Tennessee}\\
Knoxville, USA \\
hthapliyal@utk.edu}
}

\maketitle

\begin{abstract}
Peephole optimization of quantum circuits provides a method of leveraging standard circuit synthesis approaches into scalable quantum circuit optimization. One application of this technique partitions an entire circuit into a series of peepholes and produces multiple approximations of each partitioned subcircuit. A single approximation of each subcircuit is then selected to form optimized result circuits. We propose a series of improvements to the final phase of this architecture, which include the addition of error awareness and a better method of approximating the correctness of the result. We evaluated these proposed improvements on a set of benchmark circuits using the IBMQ FakeWashington simulator. The results demonstrate that our best-performing method provides an average reduction in Total Variational Distance (TVD) and Jensen-Shannon Divergence (JSD) of 18.2\% and 15.8\%, respectively, compared with the Qiskit optimizer. This also constitutes an improvement in TVD of 11.4\% and JSD of 9.0\% over existing solutions.
\end{abstract}

\section{Introduction}
Peephole optimization of quantum circuits is a very effective and scalable optimization technique which selects classically-tractable sections (peepholes) of a quantum circuit and optimizes each section. This allows optimization techniques with poor scaling, such as resynthesis, to be applied to large circuits.

Full-circuit peephole optimization methods, such as \cite{QGo,TopAS}, partition whole quantum circuits of $n$ qubits into peepholes containing at most $k$ qubits, and then perform resynthesis on each partitioned component. The resulting component circuits are then reassembled into an optimized full circuit, taking the better of each pair of subcircuits (the original or the resynthesized version). Partitioning the circuit before applying resynthesis reduces the time complexity of the process from $O(exp(n))$ to $O(exp(k))$.

\begin{figure*}
\centering
\includegraphics[trim= 0cm 0cm 0cm 0cm, width=\textwidth]{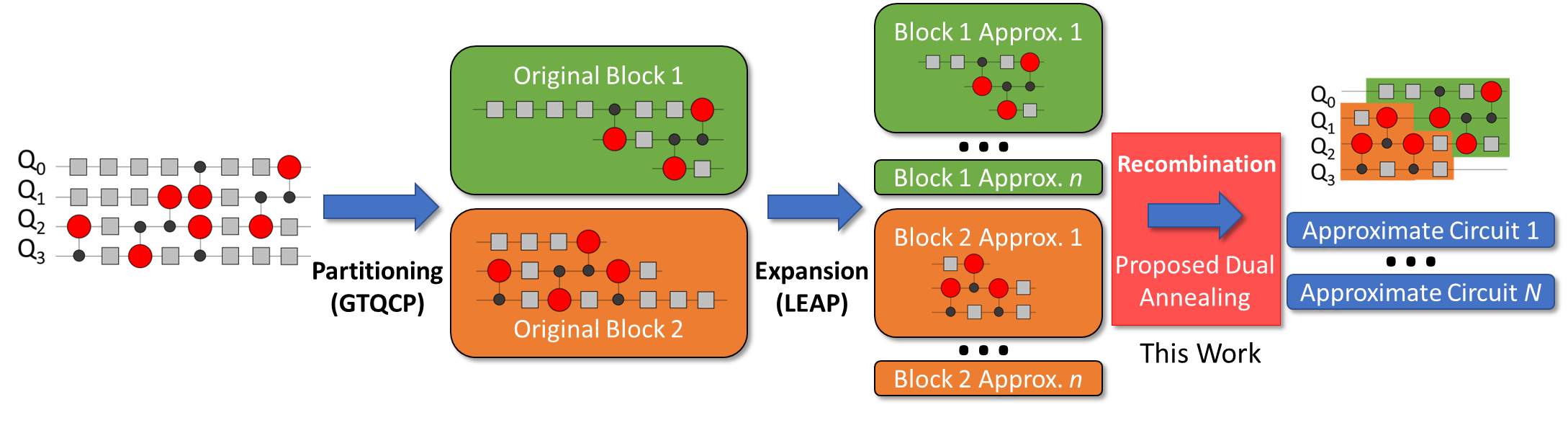}
\caption{Basic structure of the Quest framework, with three phases. Partitioning splits the circuit, expansion approximates each partition, and recombination puts approximations together to produce one or more noise resilient approximations. Recombination is the focus of this work. For the partitioning phase, we use GTQCP \cite{GTQCP}, which is a much more efficient alternative to ScanPartitioner \cite{BQSKitUpdated}. For the expansion phase, we use the modified LEAP compiler \cite{LEAP} proposed in Quest \cite{QEst}.}
\label{structure}
\end{figure*}

Quantum approximate circuit design is gaining traction due to the noise resilient qualities of these circuits \cite{NACL,quantum_assisted_compiling,QVC_noise_resilience}. An approximate variant of full circuit peephole optimization, Quest \cite{QEst}, generates multiple approximations of each partition and attempts to create a set of result circuits which can more closely match the ideal output than the original circuit when executed on noisy hardware. This is accomplished by adding an additional step to the synthesis process, which we call "recombination", shown in Figure \ref{structure}. As the recombination step is responsible for selecting partition approximations to produce optimal circuits, it has a substantial effect on the quality of the result. Thus, improving the recombination step would produce significant performance improvements.

The recombination technique employed in Quest is a dual annealer which explores the set of possible subcircuit combinations. The chosen objective function is composed of three main metrics: (1) one which ensures that the process distance (defined as the Hilbert-Schmidt inner product \cite{hilbert_schmidt_product}) between the approximation and the original circuit is within some acceptable range (by default 0.1); (2) a complexity reduction metric that reduces the number of CNOT gates, which is meant to minimize the effect of hardware error on the circuit; and (3) a differentiation metric which encourages the selection of result circuits that are different than the ones already selected.

This method produces good results for many applications, but it has some significant limitations. 1) Although limiting the approximation error of each result circuit while minimizing the number of multi-qubit gates is theoretically sound, in practice it leaves much to be desired. In addition to introducing another parameter to consider, the approach fails to consider other sources of error, such as thermal noise and interactions with the environment, which more are strongly correlated to circuit depth than CNOT count. 2) In addition, while the sum of partition process distances is proven to provide an upper bound on the overall process distance, estimating circuit performance this way does not give any consideration to the interactions between partitions. For example, it may turn out that a small error in one partition becomes much larger when propagated to the next partition, or that a large error in one partition mostly cancels out with another error in the next. 3) Iteratively selecting partitions has the effect of producing better approximations at the beginning of the process, and significantly worse approximations later on as more circuits exist to compare new circuits against.

Additionally, while this method has been shown to perform well in more favorable conditions (smaller circuits or uniform noise, fully connected hardware), testing reveals that when circuits are simulated on hardware with more complex errors and limited connectivity, performance is substantially degraded.

We propose three new recombination techniques to address these limitations, in addition to making several smaller optimizations to the original method and some changes to the original flow. To address the first limitation, we propose an error aware circuit fidelity evaluation, which combines the apparently opposed objectives of retaining circuit functionality and reducing CNOT count while also accounting for other sources of error. To address the second point, we implement a cascaded error estimation method, which considers partition pairs rather than individual partitions. This allows the method to account for the error which happens as a result of interactions between partitions. To address the third point, we implement a population-based annealing approach, which performs annealing on multiple candidates at once, and provides all candidates to the objective function for evaluation.

To evaluate the proposed techniques, we created four recombination configurations, each of which implements one or more of the proposed techniques. We have also included the recombination method used in Quest, as well as an improved variant of that method. These configurations were evaluated by mapping a series of test circuits to the IBMQ Washington computer, running them through the approximation process, optimizing the results with Qiskit \cite{Qiskit}, and simulating using the FakeWashington backend included in Qiskit. Simulation results were compared with the ideal measurement results using two output distance metrics: Total Variational Distance (TVD) and Jensen-Shannon Divergence (JSD) \cite{output_distance}. The most promising configuration, the population-based method with error awareness, achieves a reduction in TVD from the ideal result of 18.2\% and in JSD of 15.8\% when compared with the original mapped circuit. When compared with the result from the Quest method, our best method reduces TVD by 11.4\% and JSD by 9.0\%. This method also reduces multi-qubit gate count by an average of 37.1\% from the baseline and 16.9\% over the Quest method.

This paper is organized as follows: Section \ref{prop_work} describes our proposed techniques and recombination configurations. Section \ref{res} discusses how the recombiner configurations were evaluated along with results. Section \ref{conc} discusses future research direction and concludes the article.

\section{Proposed Methods}
\label{prop_work}

\begin{algorithm}
\caption{Basic structure of annealing objective function.}
\label{objective_function}
\begin{algorithmic}[1]
\State $P_{i} \gets$ Original partitions
\State $P \gets$ Partitions to evaluate
\State $S \gets$ The set of existing partitions
\State $\epsilon \gets$ Approximation threshold
\State $w \gets$ Weight parameter
\If{$P \in S$}
\State \textbf{return} $2.2$
\ElsIf{$\braket{P|P_{i}}_{HS}>\epsilon$}
\State \textbf{return} $\braket{P_{i}|P}_{HS}-\epsilon+1.1$
\Else
\State $t \gets 0$
\ForAll{$s \in S$}
\State $t \mathrel{+}= \braket{P|s}_{HS} \leq \max{(\braket{P_{i}|P}_{HS}, \braket{P_{i}|s}_{HS})}$
\EndFor
\State $t \mathrel{/}= |S|$
\State $g \gets $ \Call{CNOT}{$P$} $/$ \Call{CNOT}{$P_{i}$}
\State \textbf{return} $w \cdot g + (1-w) \cdot t$
\EndIf
\end{algorithmic}
\end{algorithm}

Our three proposed methods each rework a different part of the desired objective function. The cascaded error estimation improves the accuracy of the approximation limitation, while the error aware fidelity evaluation combines the approximation limitation and complexity reduction steps to produce an estimation of fidelity on the target hardware. Finally, the population-based annealer allows the objective function to be optimized over all result circuits at once, rather than iteratively producing single circuits, which ensures all circuits are equally affected by the differentiation metric.

In addition to the three proposed methods, we have implemented several smaller changes to the Quest objective function and implemented them in our own methods where relevant. First, we modify the approximation limitation to produce a gradient based on the amount of excess distance between the approximate and exact circuits rather than returning a constant value. This allows the annealer to explore the search space significantly more efficiently, and tends to allow access to formerly inaccessible regions of the search space. We also corrected a small but significant error in the differentiation metric which caused a tendency for results to resemble the initial exact circuit. The basic structure of this method is shown in Algorithm \ref{objective_function}, which acts as the baseline configuration for all of our approaches. In addition to these changes, memoization has been applied to the differentiation metric calculation.

\subsection{Cascaded Error Estimation}
The cascaded error estimation metric provides a more accurate estimate of the approximation error of a circuit by cascading the unitaries for adjacent partitions and calculating the process distance between that result and the same pair of partitions in the original circuit, rather than comparing individual partitions. To facilitate this calculation, we construct a graph of the partition order, where nodes represent partitions and edges represent the flow of information between them, in the form of qubits. In order to evaluate a partition, we take the average of the distances for each pairing of a partition and its immediate neighbors on the partition graph. Each edge (qubit) connecting a pair partitions proportionally increases the weight of that pair. To evaluate a circuit, we simply sum the scores of each partition composing the circuit.

\subsection{Error Aware Fidelity Evaluation}
The introduction of an error aware fidelity evaluation provides a significant structural improvement for our objective function by combining two seemingly opposing metrics and making the minimum accuracy parameter obsolete. We implement this objective function by calculating the probability density matrix of each partition in the initial circuit to use as the baseline. We then calculate the probability density matrix for each approximation running in a simulator of the target hardware without readout error. Thus, where the original method finds the process distance between the unitaries of the approximations and the exact circuit, we calculate the Frobenius norm of the difference between the ideal density matrix and one which results from the error simulation. We calculate the average of the distances of all partitions composing a circuit as the fidelity estimate, which becomes the new complexity reduction metric in place of reducing multi-qubit gates.

\subsection{Population-Based Annealing}
The Quest recombination approach performs dual annealing once for each desired result circuit, adding each result to a list of prior results. The prior results are then used in the differentiation metric to score new circuits. However, this means that the first circuit produced does not account for any other circuits, while the last circuit is expected to be differentiated from all other circuits. Thus, rather than producing a set of well-distributed approximate circuits which average to cancel out hardware error, this approach produces an initial circuit with minimal CNOT count, with subsequent circuits becoming increasingly large and vulnerable to errors. To address these concerns, we propose a population-based annealer, which performs annealing on each member of a population of candidate solutions simultaneously. This allows all solutions to be equally influenced by the differentiation metric. In order to implement this metric, we have modified several sections of an existing dual annealing implementation \cite{scipy_lib}. Namely, the main loop of the annealer now updates all solutions in each timestep, and saves a set of results when reannealing, rather individual results. We also added an argument to the objective function which contains all solutions except the solution to be evaluated, to enable the implementation of the differentiation metric. In addition to these changes, we modify the objective function to allow duplicate results, as the improved differentiation behavior should allows the annealer to decide if duplicates are desirable.

\subsection{Time Complexity}
\textbf{Cascaded Error Estimation:} The cascaded error estimation metric is significantly more complex than the baseline metric; where the baseline version simply sums the distance between the unitaries of the original partitions and the selected approximations, the cascaded metric must sum distance between the combined unitary for each pair of adjacent selected partitions and the original pairs. Given that the metrics are evaluated $e$ times and the cost of producing the unitary for a partition with $k$ qubits is $O(2^{3k})$, the time complexity of the baseline method is $O(a2^{3k} + ep)$, where $a$ is the total number of partition approximations and $p$ is the total number of partitions. The time complexity of the cascaded metric is slightly more complex. First, the combined unitary must be prepared for each adjacent pair of partitions. Given adjacent partitions $i$ and $j$ and $a_x$ approximations for partition $x$, there are $a_i \cdot a_j$ possible pairs. The worst-case size of a pair of partitions is $2k-1$ qubits, since adjacent partitions share at least one qubit. The complexity of producing the combined unitary is therefore approximately $O(2^{6k})$, with the cost of producing all unitaries for partition pair $(i,j)$ being $O((a_i \cdot a_j) 2^{6k} + ep)$. Thus, while the cascaded error approximation is able to produce a significantly better error estimation than the baseline method, this comes at the cost of higher time complexity. It should also be noted that the number of pairs of partitions and number of approximations for each partition varies greatly depending on the circuit and the behavior of the expansion step, and thus the run time for some circuits is much lower than the time complexity suggests.

\textbf{Error Aware Fidelity Evaluation:} The time complexity analysis of the error aware fidelity evaluation is relatively straightforward. The method replaces both the baseline error estimation metric and the complexity reduction metric. Given that the metrics are evaluated $e$ times, the error estimation metric has a time complexity of $O(a2^{3k} + ep)$, while the complexity reduction metric has a much smaller time complexity of $O(ep)$. Like the baseline error estimation metric, the error aware fidelity evaluation must perform a simulation of all $a$ partition approximations and sum the distance between the ideal and approximate results. However, the time complexity of density matrix simulation is slightly lower than unitary simulation, at $O(2^{2k})$, producing a time complexity of approximately $O(a2^{2k} + ep)$. Thus, the time complexity of the error aware fidelity evaluation is faster than the baseline variant, although the run time for this method is likely to be longer than the baseline method in most cases due to the overhead induced by the error simulation.

\textbf{Population-Based Annealing:} Any difference in time complexity between the baseline and Population-based approach would necessarily be caused by differences in the time complexity of the differentiation metric, as this is the only part of the objective function which accounts for other result circuits. The baseline method compares new solutions to each existing solution. Thus, when the first circuit is being generated, the differentiation metric is not performed; however, with each new circuit generated, the number of circuits to compare against increases by one. Thus, the final solution must compare with $c-1$ other circuits, where $c$ is the total number of circuits to be generated. As the differentiation metric calculates the unitary distance between all partitions in the current and existing solution for each existing solution, the time complexity for a single evaluation of the differentiation metric is $O(i \cdot p4^k))$, where p is the number of partitions, $4^k$ is the size of each partition unitary, and i is the number of existing result circuits. Thus, given that the metric is evaluated $e$ times to generate each circuit, we can calculate that the time complexity for the original differentiation metric will be $O(e[\sum_{i=0}^{c-1} i \cdot p4^k])$. However, if this result is rearranged to move the constant terms outside of the sum, it can be simplified using the formula $\sum_{i=1}^{n} i = \frac{n(n+1)}{2}$ to produce $O(e \cdot p4^k \cdot c^2)$. Unlike the basic method, the Population-based method creates all result circuits in parallel. This means that on each iteration of the annealer, each circuit is compared with all $c-1$ other circuits, giving a time complexity of $O((c-1) \cdot p4^k)$ to evaluate one circuit. Assuming, again, that each circuit is evaluated $e$ times, this produces a total time complexity of $O(e \cdot p4^k \cdot c^2)$. Thus, the time complexity of the Population-based method is the same as that of the basic method.

\subsection{Configurations}

\newcolumntype{Y}{>{\centering\arraybackslash}X}
\begin{table}
\caption{Description of the six different configurations.}
\begin{tabularx}{\columnwidth}{|>{\columncolor[gray]{0.8}}l|Y|Y|Y|Y|}
\hline
\rowcolor[rgb]{0.88,1,1}
Configuration & Basic Changes       &  Cascaded Error  &   Error Awareness & Population-Based \\
\hline
Quest         &                    &                  &                   &                  \\
Basic         &         X          &                  &                   &                  \\
Basic w/Err   &         X          &                  &         X         &                  \\
Pop.          &         X          &                  &                   &         X        \\
Pop. w/Err    &         X          &                  &         X         &         X        \\
Cascade       &         X          &        X         &                   &                  \\
\hline
\end{tabularx}
\label{configurations_info}
\end{table}

Aside from the cascaded error estimation and the error aware fidelity evaluation (which affect the same parts of the evaluation) the proposed improvements can be applied in tandem. As a result, we have produced five separate candidate configurations in addition to the \underline{Quest} configuration, which are shown in Table \ref{configurations_info}. The first of these configurations, \underline{Basic}, is our improved variant of \underline{Quest}, which is the basic structure on which our other configurations are built. The next two are \underline{Population}, which employs the population-based approach, and \underline{Cascade}, which uses the cascaded error estimation. The final two are error aware variants of the improved Quest method and the population-based approach, called \underline{Basic with Error} and \underline{Population with Error}, respectively. As in Quest, we set the $w$ (weight) parameter for all configurations to 0.5.

We made minor changes to the rest of the Quest pipeline. Notably, we use GTQCP \cite{GTQCP} for the partitioning phase rather than ScanPartitioner \cite{BQSKitUpdated}, as GTQCP has a significantly better time complexity. For the expansion step, we use the modified LEAP compiler \cite{LEAP} proposed in Quest. We also added an initial step based on the SABRE \cite{SABRE} algorithm which maps each circuit to the target hardware before partitioning.

\section{Results}
\label{res}

\begin{table}
\caption{Benchmark circuits used to evaluate recombination methods.}
\begin{tabularx}{\columnwidth}{|>{\columncolor[gray]{0.8}}X|l|X|X|}
\hline
\rowcolor[rgb]{0.88,1,1}
           Circuit & Description &  Qubit Count  &   CNOT Count \\
\hline
             Adder & Quantum adder                                &  4 &   24 \\
                   &                                              &  9 &   98 \\
\hline
               HLF & Hidden linear function                       &  5 &   14 \\
                   & circuit                                      & 10 &   56 \\
\hline
        Multiplier & Quantum multiplier                           &  5 &   20 \\
                   &                                              & 10 &  163 \\
\hline
              QAOA & Quantum approximate                          &  5 &   42 \\
                   & optimization algorithm                       & 10 &   85 \\
\hline
               QFT & Quantum Fourier transform                    &  5 &   33 \\
                   & circuit                                      & 10 &  216 \\
\hline
              TFIM & Transverse-field Ising                       &  4 &   12 \\
                   & model simulation                             &  8 &   56 \\
\hline
               VQE & Variational quantum eigensolver              &  4 &   74 \\
\hline
                XY & XY quantum Heisenberg model                  &  4 &   12 \\
                   &                                              &  8 &   56 \\
\hline
\end{tabularx}
\label{circuit_info}
\end{table}

\begin{figure*}[t]
\centering
\includegraphics[trim= 0cm 0cm 0cm 0cm, width=\textwidth]{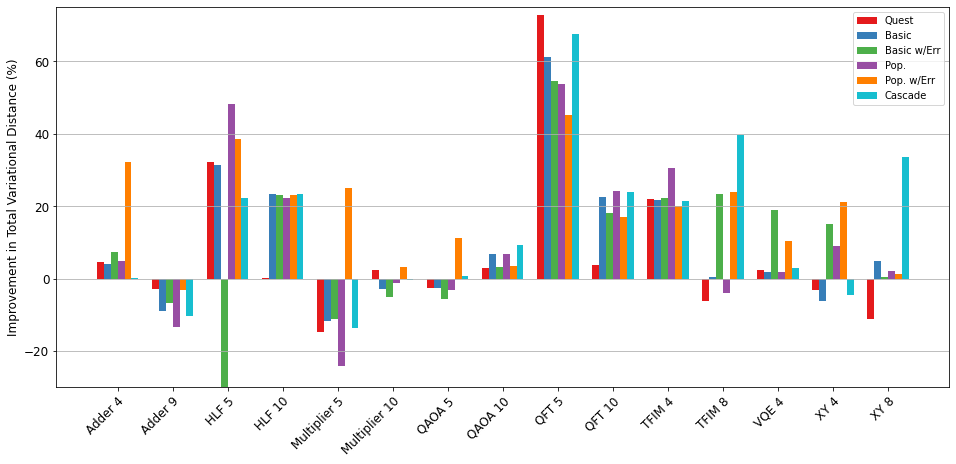}
\caption{Improvement in Total Variational Distance with respect to the optimized initial circuit across all recombination configurations for all benchmark circuits, as a percentage. Not shown is the performance of the basic method with error awareness for HLF 5, which is -92.1\%.}
\label{tvd_improvement}
\end{figure*}

\begin{figure*}[t]
\centering
\includegraphics[trim= 0cm 0cm 0cm 0cm, width=\textwidth]{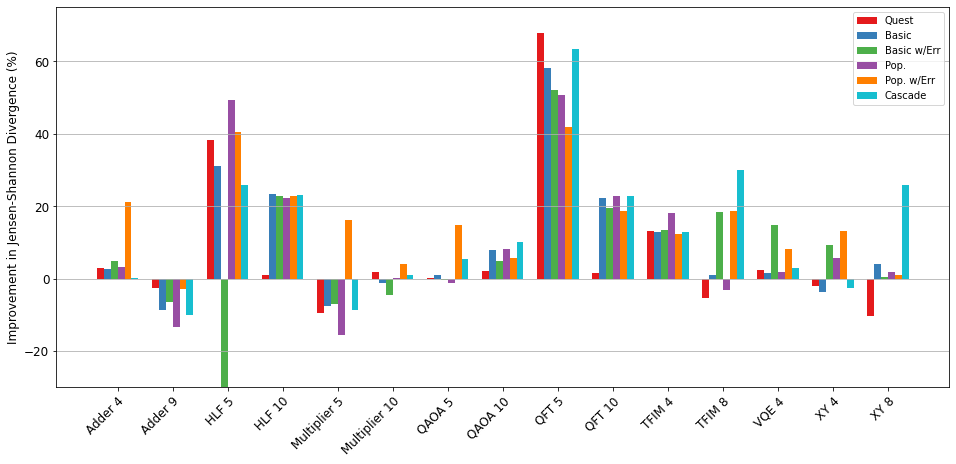}
\caption{Improvement in Jensen-Shannon Divergence with respect to the optimized initial circuit across all recombination configurations for all benchmark circuits, as a percentage. Not shown is the performance of the basic method with error awareness for HLF 5, which is -71.1\%.}
\label{jsd_improvement}
\end{figure*}

\begin{figure*}[t]
\centering
\includegraphics[trim= 0cm 0cm 0cm 0cm, width=\textwidth]{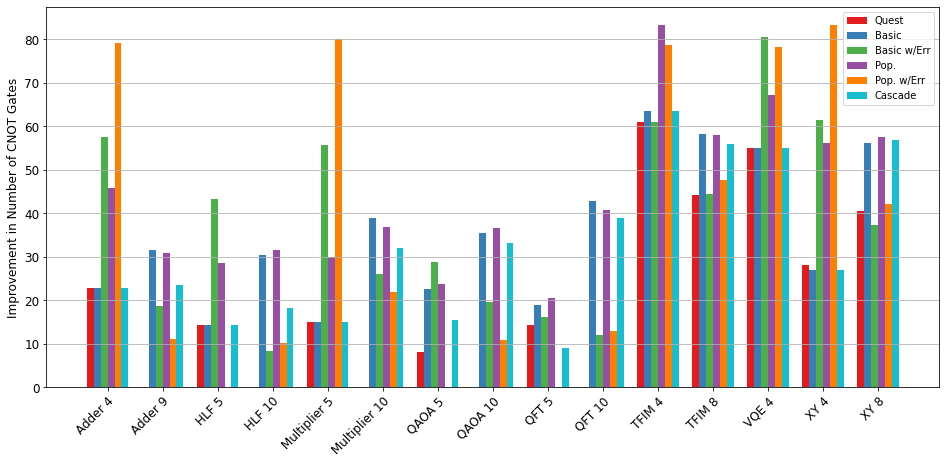}
\caption{Improvement in number of CNOT gates across all recombination configurations with respect to the optimized initial circuit for all benchmark circuits, as a percentage.}
\label{CNOT_improvement}
\end{figure*}

The six recombiners (as shown in Table \ref{configurations_info}) were implemented in the BQSKit quantum synthesis library \cite{BQSKitUpdated} and evaluated by running the optimization pipeline with each recombiner configuration for each benchmark circuit. The benchmark circuits are provided in Table \ref{circuit_info}, along with a brief description of each circuit and the qubit and CNOT gate counts of each circuit. Each benchmark circuit was mapped and partitioned with $k$ (the number of qubits per partition) fixed at $4$. The approximations produced by the subsequent expansion step were then recombined using each of the six recombiner configurations. The resulting circuits, along with the initial hardware mapped circuit, were optimized with Qiskit \cite{Qiskit} with all optimizations on, and simulated on the IBMQ FakeWashington backend with 1024 shots for each circuit. The mapped circuit was also run in an ideal simulator with optimizations disabled at 8192 shots. Performance was evaluated by calculating the Total Variational Distance and Jensen-Shannon Divergence of each combined set of circuit results in comparison with the ideal results. These results are presented in Figures \ref{tvd_improvement} and \ref{jsd_improvement}.

The results demonstrate that while the \underline{Quest} performs well on a few circuits, most notably QFT 5, performance is generally worse than the other methods tested. Similarly, both \underline{Basic} and \underline{Basic with Error} do not demonstrate particularly impressive performance for any of the benchmark circuits. While \underline{Basic with Error} performs better than the other configurations on VQE 4, it also performs significantly worse than any other configuration on HLF 5. \underline{Population} demonstrates some encouraging results, producing the best results out of all methods on HLF 5, and does not have any particularly poor benchmarks. However, the two best configurations by far are \underline{Population with Error} and \underline{Cascade}. \underline{Population with Error} gives at least some improvement on almost all test circuits, the only exception being Adder 9, which all configurations perform somewhat poorly on. However, even on Adder 9, \underline{Population with Error} is among the better performing circuits. \underline{Cascade} does not provide the best performance on every circuit, but it is among the better performing approaches for most benchmarks, and performs far better than any other circuit on the TFIM 8 and XY 8 benchmarks.

The average reduction in the number of CNOT gates for each set of results was also calculated with respect to the exact circuit, the results for which are shown in Figure \ref{CNOT_improvement}. The figure shows that all recombiners generally reduce the number of CNOT gates in the original circuit, in some cases by up to 80\%, although there is still significant variation between recombiners. For example, \underline{Quest} generally offers the lowest reduction in CNOTs, often not reducing CNOT count at all. The \underline{Basic}, \underline{Cascade} and \underline{Population} approaches generally offer similar reductions in CNOT count, with \underline{Cascade} being the lowest and \underline{Population} being the highest. The methods which stand out the most are the two error aware methods, which often produce considerably more reductions than the other methods. However, in several cases, \underline{Population with Error} actually reduces CNOT count considerably less than other configurations.

\begin{table*}
\centering
\caption{Performance improvement of recombination methods over the original circuit.}
\begin{tabularx}{\textwidth}{|>{\columncolor[gray]{0.8}}X|X|X|X|X|X|X|}
\hline
\rowcolor[rgb]{0.88,1,1}
       &       &       \multicolumn{5}{c|}{Proposed Methods}      \\
\hhline{|>{\arrayrulecolor[rgb]{0.88,1,1}}->{\arrayrulecolor{black}}|>{\arrayrulecolor[rgb]{0.88,1,1}}->{\arrayrulecolor{black}}|*5{-}}
\rowcolor[rgb]{0.88,1,1}
Metric & Quest \cite{QEst} & Basic & Basic w/Err & Pop. & Pop. w/Err & Cascade \\
\hline
 TVD & 6.8\% & 9.8\% & 4.4\% & 10.5\% & 18.2\% & 14.4\% \\
 JSD & 6.8\% & 9.7\% & 4.7\% & 10.1\% & 15.8\% & 13.5\% \\
 CNOT Reduction & 20.2\% & 35.5\% & 38.1\% & 43.2\% & 37.1\% & 32.1\% \\
\hline
\end{tabularx}
\label{perf_summary}
\end{table*}

Table \ref{perf_summary} shows the average performance improvement for each method on each benchmark circuit in terms of TVD, JSD, and CNOT count reduction. The results summary reaffirms the evaluation that the \underline{Population with Error} and \underline{Cascade} configurations perform the best, with an average improvement in TVD of 18.2\% and 14.4\%, respectively. Similar results are seen for JSD, with an average improvement of 15.8\% and 13.5\%, respectively. \underline{Quest} achieves an average reduction in TVD and JSD of only 6.8\%, giving \underline{Population with Error} an advantage of \textasciitilde{}10\% on both metrics. In terms of CNOT count reduction, the error aware methods are predictably among the better performing approaches, although the population-based approach does the best with an average improvement of 43.2\%. The \underline{Cascade} and \underline{Basic} configurations are comparable, while \underline{Quest} suffers a drop of roughly 12\% in comparison with the next closest method. The best performing methods, \underline{Population with Error} and \underline{Cascade}, show an improvement in CNOT gate count of 37.1\% and 32.1\%, respectively, compared with the original circuit. This constitutes a 16.9\% improvement for \underline{Population with Error} over \underline{Quest}.

\section{Discussion}
The extremely varied performance across most of the recombination configurations for most of the benchmarks raises a number of questions. The first concern is the relatively poor performance of the "enhanced" \underline{Quest} approach (\underline{Basic}), particularly when performance is below the original method. In these cases, the performance drop is paradoxically caused by the improvement to the part of the objective function which limits the overall error. The original objective function's exploration of the search space is so significantly limited by the faulty error limitation that in most cases, particularly on simpler circuits, only a few circuits are returned. The improved behavior does not suffer from this problem, but in several of these cases the additional circuits which are found are not of good quality. The \underline{Basic with Error} configuration suffers a similar problem compared with \underline{Population with Error}, as \underline{Population with Error} is allowed to return the same circuit more than once, where Basic with Error is not. Thus, \underline{Basic} and \underline{Basic with Error} are occasionally forced to produce poor quality circuits. Although the recombination stage includes a mechanism which terminates the process early if circuits are of poor quality, it focuses on error rather than fidelity. Thus, we suspect a better early-termination mechanism would alleviate most of these performance problems.

The cause for the poor performance of the \underline{Cascade} configuration on several of the smaller circuits is likely due to the cascade metric breaking down and not providing any significant benefit for circuits with few partitions. In these cases the cascade metric should behave similarly to the metric used the \underline{Quest} and \underline{Basic} configurations, which is indeed demonstrated by the results. As this is a limitation of the cascade approach, it is not possible to resolve this issue by modifying the metric. However, this behavior likely indicates that a hybrid between the cascade approach and one of the other proposed improvements could yield significant performance benefits.

Several circuits see poor performance across most or all recombiner configurations, notably Adder 9, Multiplier 10, and QAOA 10. Unlike some other challenging circuits where at least one method sees good performance, the poor performance in these cases seems to be caused by low quality approximations produced by the expansion step. For all three of these circuits, the modified approximate compiler which implements the expansion step often fails to produce a single approximation with fewer CNOT gates than the original and an error of less than 0.2. This explains why these benchmarks see little or no improvement over the original, as often the only viable circuits are largely composed of the original exact partitions. Even more problematically, the compiler also frequently produces approximations with fewer CNOT gates than the original partition but many more single-qubit gates. Although multi-qubit gates generally have a higher error rate than single-qubit gates, the difference is not large enough to warrant such a drastic trade-off. This explains why the \underline{Population with Error} approach generally performs better than the other configurations on these benchmarks, as error awareness and the ability to repeat result circuits makes the approach more adaptable when presented with few useful approximations. This issue could be resolved by selecting a different expansion method or investigating and resolving the performance issues experienced by the current one.

\section{Conclusion}
\label{conc}
Full-circuit peephole optimization provides an interesting method for producing error resilient approximations of a given circuit which do not deviate too significantly from the exact output. However, limitations in the recombination step of existing methods must be addressed before these methods can be applied to larger quantum circuits. Notably, the recombination method proposed in Quest \cite{QEst} has several shortcomings, including difficulty balancing correctness and complexity reduction, difficulty propagating approximation errors through circuits, sub-optimal differentiation metrics, and poor performance on circuits which have been mapped to restricted hardware. We address each of these problems by proposing changes to the recombination objective function, with the best performing set of changes seeing an 18.2\% decrease in Total Variational Distance (TVD) and a 15.8\% decrease in Jensen-Shannon Divergence (JSD) over the Qiskit-optimized exact circuit. This corresponds to an 11.4\% and a 9.0\% improvement over Quest in TVD and JSD, respectively. 

Although the proposed methods provide good improvements over Quest with no significant impact on scalability, there is still much room for improvement. While the \underline{Population with Error} and \underline{Cascade} configurations perform well, their performance is still quite poor on several circuits, and very inconsistent. We suspect that an error aware method with cascaded evaluation might perform better, but implementation of such a method is complicated by the fact that the cascade metric operates on circuit unitaries, while the error aware metrics estimate error using the probability density matrices produced by each circuit. Finally, we suspect that poor performance on the Adder 9, Multiplier 10, and QAOA 10 benchmarks may be due to poor approximation quality, meaning that improvements in the approximate circuit generation step may produce significant performance improvements.

The proposed work presents a significant improvement to the existing Quest approach, improving the framework's ability to adapt to a variety of noise sources and levels, and the quality of approximate circuits for more complex quantum algorithms. This enhances real-world usability, enabling more useful results to be extracted from NISQ computers due to improved noise resilience.

\section*{Acknowledgments}
This research used resources of the Oak Ridge Leadership Computing Facility, which is a DOE Office of Science User Facility supported under Contract DE-AC05-00OR22725.
Also, this material is based upon work supported by the National Science Foundation Graduate Research Fellowship under Grant No. 1938092.

\section*{References}

\printbibliography[heading=none]

@misc{BQSKitUpdated,
  author = {Berkeley National Laboratory},
  title = {BQSKit},
  year = {2022},
  publisher = {GitHub},
  journal = {GitHub repository},
  howpublished = {\url{https://github.com/BQSKit/bqskit}},
  commit = {3330d7bb917f4f831dfc17c8088b87bf25b80144}
}

@inproceedings{QGo,
  title={Reoptimization of quantum circuits via hierarchical synthesis},
  author={Wu, Xin-Chuan and Davis, Marc Grau and Chong, Frederic T and Iancu, Costin},
  booktitle={2021 International Conference on Rebooting Computing (ICRC)},
  pages={35--46},
  year={2021},
  organization={IEEE}
}

@inproceedings{QEst,
  title={Quest: systematically approximating quantum circuits for higher output fidelity},
  author={Patel, Tirthak and Younis, Ed and Iancu, Costin and de Jong, Wibe and Tiwari, Devesh},
  booktitle={Proceedings of the 27th ACM International Conference on Architectural Support for Programming Languages and Operating Systems},
  pages={514--528},
  year={2022}
}

@article{LEAP,
  title={Leap: Scaling numerical optimization based synthesis using an incremental approach},
  author={Smith, Ethan and Davis, Marc Grau and Larson, Jeffrey and Younis, Ed and Oftelie, Lindsay Bassman and Lavrijsen, Wim and Iancu, Costin},
  journal={ACM Transactions on Quantum Computing},
  volume={4},
  number={1},
  pages={1--23},
  year={2023},
  publisher={ACM New York, NY}
}

@Misc{scipy_lib,
  author =    {Eric Jones and Travis Oliphant and Pearu Peterson and others},
  title =     {{SciPy}: Open source scientific tools for {Python}},
  year =      {2001--},
  url = "http://www.scipy.org/"
}

@inproceedings{TopAS,
  title={Wide quantum circuit optimization with topology aware synthesis},
  author={Weiden, Mathias and Kalloor, Justin and Kubiatowicz, John and Younis, Ed and Iancu, Costin},
  booktitle={2022 IEEE/ACM Third International Workshop on Quantum Computing Software (QCS)},
  pages={1--11},
  year={2022},
  organization={IEEE}
}

@inproceedings{GTQCP,
  title={GTQCP: Greedy Topology-Aware Quantum Circuit Partitioning},
  author={Clark, Joseph and Humble, Travis S and Thapliyal, Himanshu},
  booktitle={2023 IEEE International Conference on Quantum Computing and Engineering (QCE)},
  volume={1},
  pages={739--744},
  year={2023},
  organization={IEEE}
}

@misc{Qiskit,
    author = {{Qiskit contributors}},
    title = {Qiskit: An Open-source Framework for Quantum Computing},
    year = {2023},
    doi = {10.5281/zenodo.2573505}
}

@article{NACL,
  title={Machine learning of noise-resilient quantum circuits},
  author={Cincio, Lukasz and Rudinger, Kenneth and Sarovar, Mohan and Coles, Patrick J},
  journal={PRX Quantum},
  volume={2},
  number={1},
  pages={010324},
  year={2021},
  publisher={APS}
}

@article{output_distance,
  title={On measures of entropy and information},
  author={Crooks, Gavin E},
  journal={Tech. Note},
  volume={9},
  number={4},
  year={2017}
}

@inproceedings{SABRE,
  title={Tackling the qubit mapping problem for NISQ-era quantum devices},
  author={Li, Gushu and Ding, Yufei and Xie, Yuan},
  booktitle={Proceedings of the Twenty-Fourth International Conference on Architectural Support for Programming Languages and Operating Systems},
  pages={1001--1014},
  year={2019}
}

@article{quantum_assisted_compiling,
  title={Quantum-assisted quantum compiling},
  author={Khatri, Sumeet and LaRose, Ryan and Poremba, Alexander and Cincio, Lukasz and Sornborger, Andrew T and Coles, Patrick J},
  journal={Quantum},
  volume={3},
  pages={140},
  year={2019},
  publisher={Verein zur F{\"o}rderung des Open Access Publizierens in den Quantenwissenschaften}
}

@article{QVC_noise_resilience,
  title={Noise resilience of variational quantum compiling},
  author={Sharma, Kunal and Khatri, Sumeet and Cerezo, Marco and Coles, Patrick J},
  journal={New Journal of Physics},
  volume={22},
  number={4},
  pages={043006},
  year={2020},
  publisher={IOP Publishing}
}

@article{hilbert_schmidt_product,
  title={An alternative quantum fidelity for mixed states of qudits},
  author={Wang, Xiaoguang and Yu, Chang-Shui and Yi, XX},
  journal={Physics Letters A},
  volume={373},
  number={1},
  pages={58--60},
  year={2008},
  publisher={Elsevier}
}

\end{document}